# Detection of Light Sleep Periods Using an Accelerometer Based Alarm System

Egemen Turkyilmaz, Alper Akgul, Erkan Bostanci and Mehmet Serdar Güzel
Ankara University Computer Engineering Department SAAT Lab
Ankara Turkey
egmntrkylmz@gmail.com, alperakgul1996@gmail.com, ebostanci@ankara.edu.tr, mguzel@ankara.edu.tr

*Abstract*—Light sleep is a sleeping period which occurs within each hour during the sleep. This is the period when people are closest to awakening. With this being the case people tend to move more frequently and aggressively during these periods. The characteristics of sleeping stages, detection of light sleep periods and analysis of light sleep periods were clarified. The sleeping patterns of different subjects were analyzed. In this paper the most suitable moment for waking a person up will be described. The detection of this moment and the development process of a system dedicated to this purpose will be explained, and also some experimental results that are acquired via different tests will be shared and analyzed.

## I. INTRODUCTION

In daily life most of the people needs an alarm system because they might have to wake up at a specific hour of the day. However, it may not be convenient to wake up at a random moment since sleep has a nature of continuous loops. It was observed that when people wake up in deep sleep stages, they usually have difficulties in awakening and encounter mental focus problems that endure for a certain period of time. There are 2 fundamental stages of sleep. These are Rapid Eye Movement (REM) and Non-Rapid Eye Movement (NREM) stages. All of the human-beings have these stages in their sleep. There are periods within the stages of sleep where people can wake up more easily. According to these researches, The NREM stage contains light sleep periods. As mentioned above, if the light sleep period is detected and the alarm is sounded during this period, it will be easier to wake up and feel more intense during the day. The main subject of this article is the development of an alarm system which aims to provide a solution to that issue.

The necessary literature reviews about the sleep stages and alarm systems, were made before the work on the subject started. The duration of sleep cycle segments differentiate throughout the maturation of the humans. Based on some researches, an adult human's REM sleep takes approximately 20% of total sleep time as of a newborn's 50% [1, 2]. In the time of REM sleep, high brain activity is detected. On the contrary, there is no excessive body movement. Also dreaming takes place in this stage [3, 4].

On the other hand NREM is the period that includes the light sleep. In this period people are in such a state that is between asleep and awakened and also during this interval an increase is observed on body movements [5]. Light sleep stage is indicated by these movement patterns. Fig. 1 depicts the time spent on REM and NREM stages during the sleep.

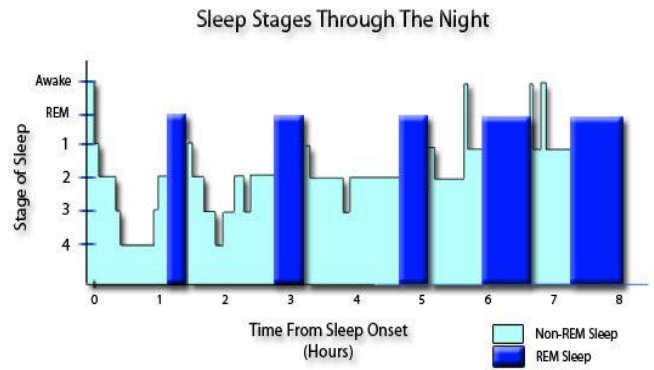

Fig. 1. Sleep stages through the night [6]

People can wake up more easily and be more energetic and strong throughout the day when they are awakened during the light sleep stage of NREM period. There are several studies and productions related to this subject. One being a mobile application called "Sleep Cycle Alarm" which uses the internal accelerometer and microphone of the device it is installed. This application works when the phone is placed somewhere nearby the sleeper. It is affected by the vibrations of the surface where the phone is placed on [7]. And also there are three patented studies that we encountered during our researches [8−10]. But these studies don't constitute an impediment to work on this topic and try to develop some other solutions.

This paper analyzes stages of the sleep and their relations with body movements and focuses on an alarm system that detects the optimal moment to wake a person up based on the change in acceleration.

The rest of the paper structured as follows. In Section II the environment of the alarm system is explained, followed by the specifications of the hardware used in the system. Section III addresses the development process of the necessary programs constructed for the system. In Section IV, the results acquired out of the experiments are shared. Finally the paper is concluded in Section V.

## II. SYSTEM ARCHITECTURE

In order to gather the necessary data, to analyze the sleep stages some hardware presence is required. In this section, the detailed features of the equipment used will be discussed. Also, the purpose of these equipments will be covered. To implement a system that meets these requirements an accelerometer and a platform to process the data is used. In this section, the hardware used will be explained in detail.

### A. Raspberry Pi

Raspberry Pi is a low-cost, portable and modular computer that plugs into a display monitor. Most types of it are capable to do most of the work that a desktop computer can do. In this research as shown in the Fig. 2, a "Raspberry Pi 2 Model B" is used. This particular device has a 900 MHz quad-core ARM Cortex-A7 CPU (Central Processing Unit) and 1GB of RAM along with 4 USB 2.0 ports, 40 GPIO (General Purpose Input Output) pins, a Full HDMI port, an Ethernet port, a micro SD card slot and a VideoCore IV 3D graphics core. The IMU is plugged into one of the USB 2.0 ports, a display monitor into the Full HDMI port, the passive buzzer into the GPIO pins. The operating system is installed on an SD card and it is inserted into the SD Card slot located on the Raspberry Pi [11,12].

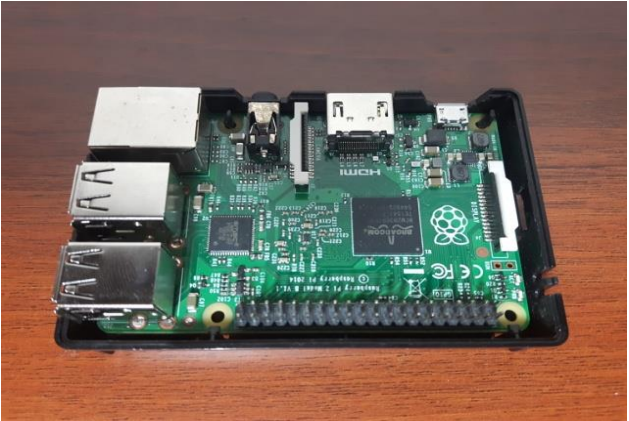

Fig. 2. Raspberry Pi 2 Model B

### B. Accelerometer

As demonstrated in Fig. 3, a "PhidgetsSpatial 1056 3/3/3" model IMU is used to measure acceleration values. This piece of hardware also includes gyroscope and magnetometer chips. This IMU provides static and dynamic acceleration, magnetic field and angular rotation measurements in 3 axes. "PhidgetsSpatial 1056 3/3/3" makes it possible to measure real-life motion in real time. With its precise voltage supply filtering, it guarantees low noise and correct sensor operation. But this paper focuses especially on IMU's static and dynamic acceleration measurements. The accelerometer can sense a minimum 228μg (alternatively: 2.2 mm/$s^2$) of change in acceleration. And it can measure a maximum acceleration of ±5g (alternatively: 49m/$s^2$). The IMU has a minimum sampling speed of 1 samples/s (samples per second) and a maximum sampling speed of 250 samples/s [13].

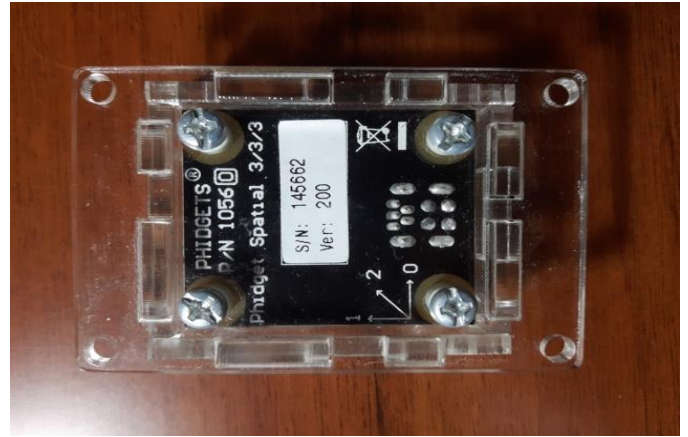

Fig. 3. PhidgetSpatial 3/3/3

### C. Passive Buzzer

A passive buzzer is a device that produces different types of sounds depending on the signals received by its pins. It is a low-cost device and can be replaced quite easily by another one. It requires a computer program to generate the desired tunes. Two jumper cables are needed to plug in the buzzer into a Raspberry Pi. In the system shown in Fig. 3, the buzzer is plugged into $11^{th}$ and $14^{th}$ GPIO pins of the Raspberry Pi. According to Raspberry Pi 2 Model B's GPIO pinout, $11^{th}$ pin is used for sending the signals from the program to the buzzer and $14^{th}$ pin is used for ground. A passive buzzer with two jumper cables is shown in Fig. 4.

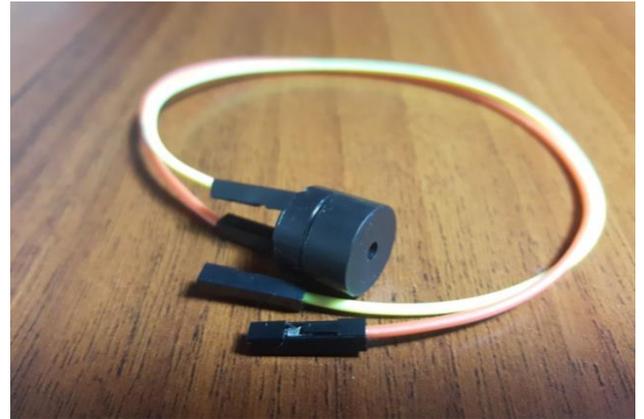

Fig. 4. Passive buzzer with two jumper cables

### D. Complete System

As the system requires an accelerometer a "PhidgetSpatial 1056 3/3/3" model IMU (inertial measurement unit) is used. This hardware is operated on a "Raspberry Pi" running a "Raspbian" operating system with necessary libraries and drivers installed. The IMU is connected to a USB 2.0 port on the Raspberry Pi as displayed in Fig. 5.

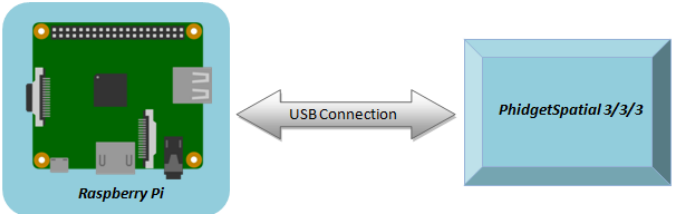

Fig. 5. Working concept of the IMU

After the installations are completed the IMU is attached to the subject's body. Then according to given sleep time as the input, the IMU starts measuring acceleration based on body movements. This process takes as long as the indicated sleep time. When the timer enters the last period of sleep time, the program seeks a specific condition to sound an alarm. This alarm is sounded by a passive buzzer plugged into the Raspberry Pi via jumper cables. As the buzzer is a passive one, it needs a dedicated program to play the desired tune. The complete system is shown in Fig. 6.

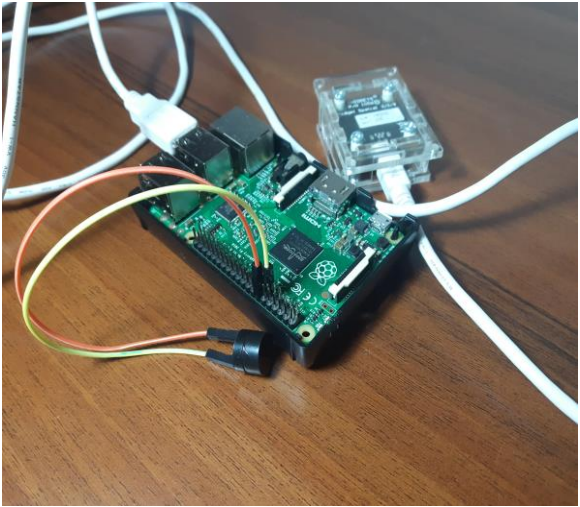

Fig. 6. The complete system

### III. SOFTWARE DEVELOPMENT

During the software development process, various libraries and programming languages are used. "WiringPi" and "Phidget22" libraries have been involved and used with C programming language. This section will mainly focus on the development process of the programs which are responsible from handling the alarm system. The software developed mainly consists of three separate program files. One being the measurement program, another handling the output files and the last one generates a tune for the alarm.

*A. Development Platform*

All of the programming in this research is done in C programming language. This programming language is chosen because it is supported by the IMU's libraries and drivers, it is capable of running on different types of CPU architectures and C programming language provides flexibility when used on embedded systems. Also on embedded systems memory and storage capacities may often be limited. C is an efficient and proper programming language to work on such kind of systems.

*B. Used Libraries*

A C library called "libphidget22" is required and used to operate the IMU. This library has an accelerometer class named "PhidgetAccelerometer" which is used to gather acceleration data from accelerometer boards of the IMU.

Through "PhidgetAccelerometer_create()" function an instance of a Phidget channel is created. It takes the reference of the channel handle and returns a "PhidgetReturnCode". All the features of the accelerometer are accessed through this channel.

After creating the channel an event which is named "Phidget_setOnAttachHandler()" assigns a handler to be called when the IMU is attached to the system. "Phidget_openWaitForAttachment" function opens the Phidget channel by "Channel" PhidgetHandle and a timeout value in milliseconds. This function blocks the access to the device until the channel is opened or a timeout occurs. To measure the change in acceleration in real-time "PhidgetAccelerometer_setOnAccelerationChangeHandler()" event is used. This event is raised continuously throughout the run time and it uses a callback function to constantly detect the changes in acceleration values. This callback function will be explained in detail later in this paper. When the device is detached an event called "Phidget_setOnDetachHandler()" is raised. Then before the program is terminated the created channel is closed by "Phidget_close()" function to make the channel ready to for another use [13].

"WiringPi" is a C library which grants access to GPIO pins in all Raspberry Pi models. With it being written in C it can be easily used while working on C programming language. This particular library includes a command-line utility named "gpio" which serves as a pin programming interface. It is also useful to read and write the GPIO pins from shell scripts [14]. With the aid of "WiringPi" library, intended melodies can be played from the buzzer.

*C. Creating the Alarm System*

Once the alarm is set, the IMU starts measuring the acceleration of the sleeper 3 or 4 times second. This rate helps to reduce the memory space needed and still gives satisfactorily detailed results. In each time period the maximum change in acceleration at any time is calculated and stored in an array. Depending on these values, a maximum and a minimum threshold value are determined until the last period of the sleep time. While measurements are made continuously in the last period, also the measured values are compared with the predetermined threshold values. This comparison is made according to (1) [15].

$$S_{stage} = \begin{cases} S_{NREM}, & \text{if } T_{min} \leq A_i \leq T_{max}, \\ S_{REM}, & \text{otherwise}, \end{cases} \quad (1)$$

If a value is measured between predetermined threshold values, then the alarm is sounded. Otherwise, the alarm is sounded at the end of the sleep time. Fig. 7, shows the flowchart of the alarm system.

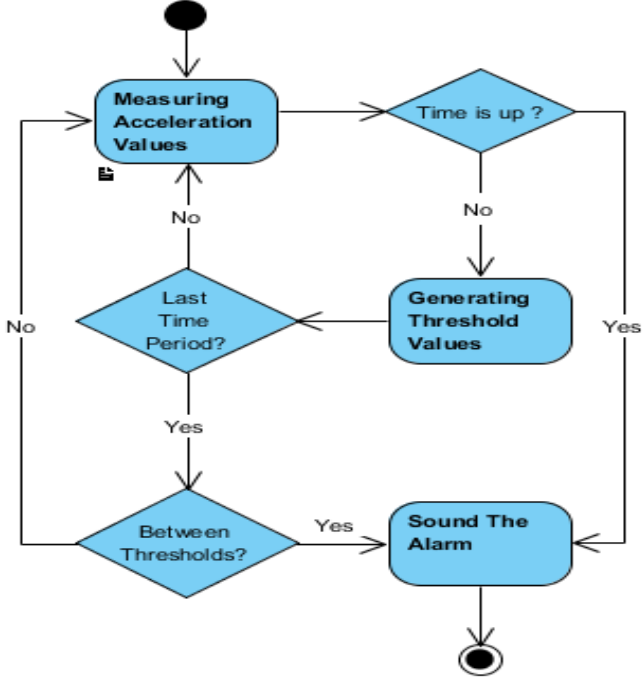

Fig. 7. Program flowchart

### D. Normalizing the Acceleration Values

Normalization process is done to gather more precise acceleration values. Each acceleration value generates a 3 dimensional vector. To normalize the acceleration values, the Euclidean length of the vector is calculated and each measurement is divided by this particular value. Thanks to this operation the noise level of the measurements is reduced and all the measured values of each axes are taken into [-1 , 1] range. Calculation of the Euclidean distance in 3 dimensions is formulated in (2).

$$dist[(x,y,z),(a,b,c)] = \sqrt{(x-a)^2 + (y-b)^2 + (z-c)^2} \quad (2)$$

$$\begin{cases} x, & acc_x / dist[(x,y,z),(a,b,c)] \\ y, & acc_y / dist[(x,y,z),(a,b,c)] \\ z, & acc_z / dist[(x,y,z),(a,b,c)] \end{cases} \quad (3)$$

Normalization in completed as shown in (3), and based on this normalized values the Manhattan distance between back to back measurements is calculated as shown in (4). These Manhattan distance values are used as input parameters to sound the alarm.

$$Md(Acc_i) = d(x) + d(y) + d(z) \quad \begin{cases} d(x), & |x_{i+1} - x_i| \\ d(y), & |y_{i+1} - y_i| \\ d(z), & |z_{i+1} - z_i| \end{cases} \quad (4)$$

## IV. EXPERIMENTAL RESULTS

During the development a large amount of short-term tests are made. The aim of these short-term tests was testing whether the system worked correctly or not. Such as observing noising levels on acceleration values, controlling the timer states, checking if the buzzer worked properly and comparing the real time measurements with stored acceleration values.

After fixing the issues encountered during the short-term tests, some different long-term tests were made. Generally these long-term tests cover about 8 hours of sleep time. The outputs of these tests are written into text files and for each hour of sleep time, separate graphic charts are drawn out of them. For all the graphs drawn in this section, the x-axis represents the time in seconds and the y-axis represents the change on acceleration in g. Fig. 8 serves an example.

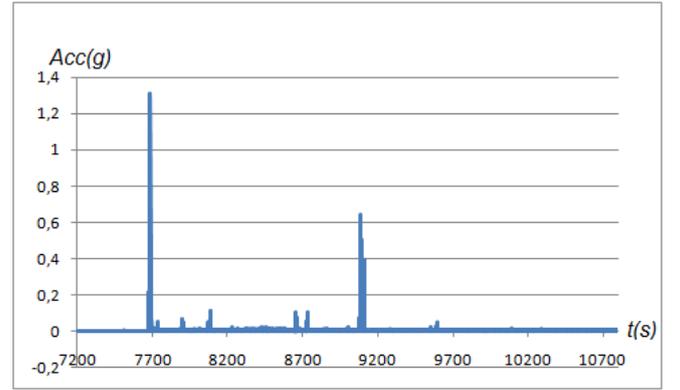

Fig. 8. Graphic chart of an hour in the sleep time

On Fig. 8, the x-axis represents time in seconds and the y-axis represents the change on acceleration in g. At the maximum peak point on the graph the subject passes to NREM period. Most of the time, the graphic charts have similar patterns when the subject is in sleep.

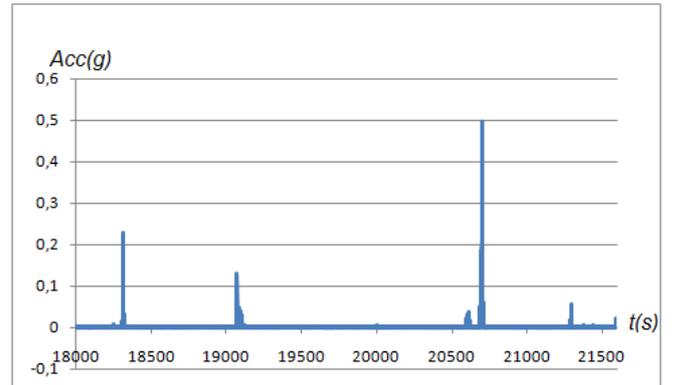

Fig. 9. Time period of minimum threshold

The minimum threshold value for this test was determined in the interval between 5[th] and 6[th] hours as 0.497g. This can be clearly seen in Fig. 9. At each hour interval the maximum change in acceleration is set as the maximum threshold value but this is updated whenever a greater value is measured.

Among those maximum acceleration changes the smallest one is set as minimum threshold value.

In Fig. 10, the interval between 4th and 5th hours is graphed. The maximum threshold value was determined as 1.662g in this interval.

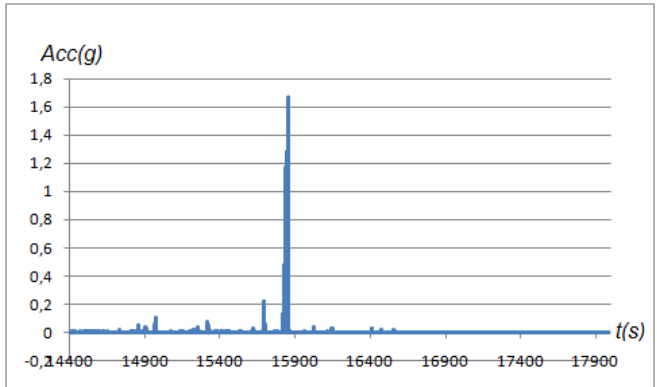

Fig. 10. Time period of maximum threshold

Due to the provision of $S_{NREM}$ condition according to (1), in the interval between 6th and 7th hours, the measurements are stopped. Therefore the alarm sounded when 1.016g was measured. Fig. 11 depicts the awakening moment of the subject in this particular test and it can be seen that from the moment when the acceleration peaked, no more acceleration values are measured. This indicates that the alarm was sounded at this point.

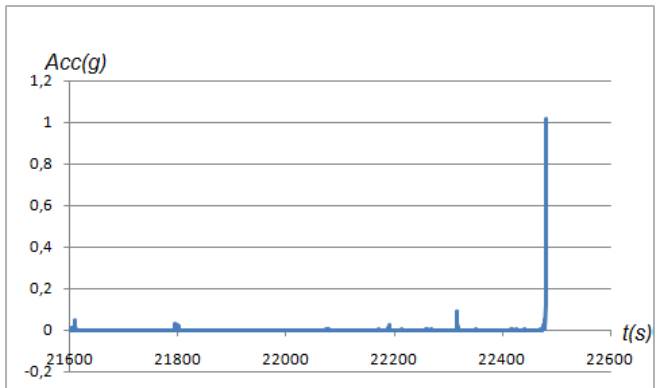

Fig. 11. Awakening moment in the last period

V. CONCLUSION

Awakening process of a human-being may prolong if they are awakened within a wrong timing. In the journey to get over this problem a good amount of experiments were conducted and many advantages of waking up within the correct timing has been observed in each particular experiment.

The tests were run on four different subjects and out of each test similar results were obtained. The result acquired belongs to one of them.

The optimal wake-up time within the last one hour varied in different tests. In the end there is a peak moment at each hour in every test. But the exact time of the NREM period can differ between the tests

As a result, a system that detects the light sleep period in the NREM stage depending on the acceleration values of the subject was built and programmed to sound the alarm when the transition to the light sleep period occurred within the desired time interval.

In order to make the system more stable, the program which operates the system can be supported by several deep learning and data mining algorithms. Long term analysis can be done according to the sleeping patterns of each particular user. Even to go further some heart-rate measuring sensors can be added on and combined with the accelerometer. In order to make the system more user-friendly, it can be integrated on a single circuit board and can be made wireless and wearable.